\documentclass{aastex63}

\usepackage{amsmath}
\usepackage{amsfonts}
\usepackage{bm}
\usepackage{graphicx}
\usepackage{subfigure}
\usepackage{threeparttable}
\usepackage{hyperref}
\usepackage{lineno}
\usepackage{soul} 

\received{Oct 6, 2023}
\revised{xxxx}
\accepted{xxxx}

\submitjournal{ApJ}

\shorttitle{The cumulation of debris clouds}
\shortauthors{Huang et al.}

\begin{document}

\title{The cumulation of debris clouds around a fast-rotating asteroid}

\correspondingauthor{Yang Yu}
\email{yuyang.thu@gmail.com, yu.yang@buaa.edu.cn}

\author{Chenyang Huang}
\affiliation{Beihang University \\
School of Aeronautic Science \& Engineering, Beijing 100191, China}

\author[0000-0001-9329-7015]{Yang Yu}
\affiliation{Beihang University \\
School of Aeronautic Science \& Engineering, Beijing 100191, China}

\author{Zhijun Song}
\affiliation{Beihang University \\
School of Aeronautic Science \& Engineering, Beijing 100191, China}

\author{Bin Cheng}
\affiliation{Tsinghua University \\
School of Aerospace Engineering, Beijing 100084, China}

\author{Patrick Michel}
\affiliation{University of Nice Sophia Antipolis \\
CNRS, Observatoire de la C\^ote d'Azur, B.P. 4229, F-06304 Nice Cedex 4, France}

\author{Hexi Baoyin}
\affiliation{Tsinghua University \\
School of Aerospace Engineering, Beijing 100084, China}

\begin{abstract}

The rotational mass loss has been realized to be a prevalent mechanism to produce low-speed debris near the asteroid, and the size composition of the asteroid's surface regolith has been closely measured by in situ explorations. However, the full-scale evolution of the shedding debris has not been examined using the observed particle sizes, which may hold vital clues to the initial growth of an asteroid moonlet, and help us to understand the general mechanisms that dominate the formation of asteroid systems. This paper presented our study on the cumulative evolution of the debris cloud formed by a rotationally unstable asteroid. A semi-analytical model is developed to characterize the spatial-temporal evolution of the debris cloud posterior to a shedding event. Large-scale DEM simulations are performed to quantify the clustering behavior of the debris particles in the mechanical environment near the asteroid. As a result, we found the cumulation of a steady debris cloud is dominated by large pieces of debris, and the shedding particles follow a common migration trend, which fundamentally determines the mass distribution of the debris cloud. For the accretion analysis, we sketched the life cycle of a debris cluster, and showed its dependency on particle size. The DEM simulations adopt physical parameters estimated from observations and asteroid missions. The results confirm porous fluffy cluster structures can form shortly after a shedding event with magnitudes as the observed shedding activities. Measurements to these structures show they possess certain strength and adsorption capacity to collisions from dissociative debris particles.

\end{abstract}


\section{Introduction} \label{sec:intro}

Various observational evidences show that a fast rotating asteroid, even without reaching the critical spin limit, can be bounded by dust/debris that is originated from the asteroid surface \citep{landj2013,hsieh2014,jewitt2014,braga2014,kleyna2019}. \citet{jewitt2015} listed $18$ currently known active asteroids with new observations for the interests of tracing their separate origins. Several physical processes are known capable of causing the emergence of debris cloud around asteroids, including rotational shedding-off, meteorite bombardments, thermal disintegration, ice sublimation, gardening effects, etc. \citep{jewitt2015}. Among these processes, the rotation is a common mechanism to drive the activity in these asteroids, which could play a dominant role for those near the critical spin limits (e.g., $6478$ Gault) or assist to destabilise the surface for those with lower spin rates (e.g. $3200$ Phaethon) \citep{pandh2000,kleyna2019,landj2013}. Besides the generation mechanism, the rotation speed of an asteroid can also have a strong influence on the subsequent evolution of the ejected particles, i.e., shaping the dynamical environment in collaboration with the non-axisymmetric gravitational field \citep{sch2015}. \citet{yu2018} developed a physical model of a top-shaped asteroid, $65803$ Didymos A, based on radar observations, showing the asymmetries in shape can produce entirely different post-shedding dynamics when the rotational period varies from $2$-$3$ hr. \citet{sicardy2019} present that modest topographic features of $10199$ Chariklo and $136108$ Haumea have essential effect on the secular stability of their rings, i.e., faster rotators reconcile the $1/2$ resonance with the Roche limit in the initial collisional disk and are beneficial to rings relatively far away from the bodies.

A number of dynamics studies focus on the ballistic motion of debris particles. \citet{richard2011} developed an excavation flow properties model that tracks the sampled ejecta particles from an impact crater, and applied this model to mimic the evolution of ejecta from cratering processes on the components of the binary asteroid $1999$KW$4$ \citep{randt2015}. Comparable models was developed by \citet{geissler1996}, which was used to track the motion of ejecta from the crater Azzurra on $433$ Ida, and by \citet{yu2017}, to track the post-impact evolution of the debris cloud in a binary asteroid system. \citet{fahnestock2014} studied the ejecta dynamics during a small-scale artificial kinetic impact onto the asteroid $101955$ Bennu, the target of asteroid OSIRIS-REx mission. Interestingly, Bennu was later identified as an active asteroid by the mission. \citet{lauretta2019} measured the events of the particle ejection from the surface of Bennu and reproduced their orbits using point-mass gravity model, which helped in estimating their source location and initial velocities. Another major topic of dynamics is the ultimate fates of the debris particles moving around asteroids. \citet{fahnestock2009} showed that the material originated from the primary of a binary asteroid can reaccrete and cause an angular momentum transport. \citet{yu2019} in their studies of the expansion rate of shedding debris, confirmed the cumulated mass may cause a spiralling-in motion of the secondary. \citet{harris2009} studied $1999$KW$4$ and showed the equatorial regolith will lift off and redeposit because of the tide from secondary, causing an effect named as tidal saltation. \citet{jands2011} built a macroscopic model showing the splitting binary components could end up with different kinds of synchronous configurations.

Many details of the debris cloud evolving around asteroids are not yet fully understood. \citet{walsh2008,walsh2012} applied the gravitational $N$-body methodology to model the YORP-driven formation of a binary asteroid system, in which they reproduced the $1999$KW$4$ system in multiple topographic features. However, as limited to the computing capacity, particles used in these simulations are fictitiously large (each in $\sim 100$ m), much larger than the average size range as revealed by the images of asteroids' surfaces \citep{miyamoto2007,walsh2019,michikami2019}. Knowledge on the cumulative effects of the debris cloud around fast-rotating asteroids is still lacking, which could be crucial because these materials provides the initial blocks that composed the new asteroid system, and the growth of grains is sensitive to the kinetic conditions of the cloud \citep{johansen2014}. In this study, we developed an informative model to measure the full scale evolution of the debris cloud shed from a rotationally unstable asteroid. Section~\ref{sec:mdl} describes the method, equations, and numerical settings of our calculation and simulation. Section~\ref{sec:result} presents the results from both the semi-analytical model (for the macroscopic propagation of the debris cloud) and the DEM model (for a detailed look into the accretion of debris particles). Section~\ref{sec:concl} summarizes the main results and conclusions of this study. 

\section{Model} \label{sec:mdl}

\subsection{The mechanics of surface shedding} \label{sec:surfdyn}

\begin{figure}[h!]
	\centering
	\includegraphics[width=0.75\textwidth] {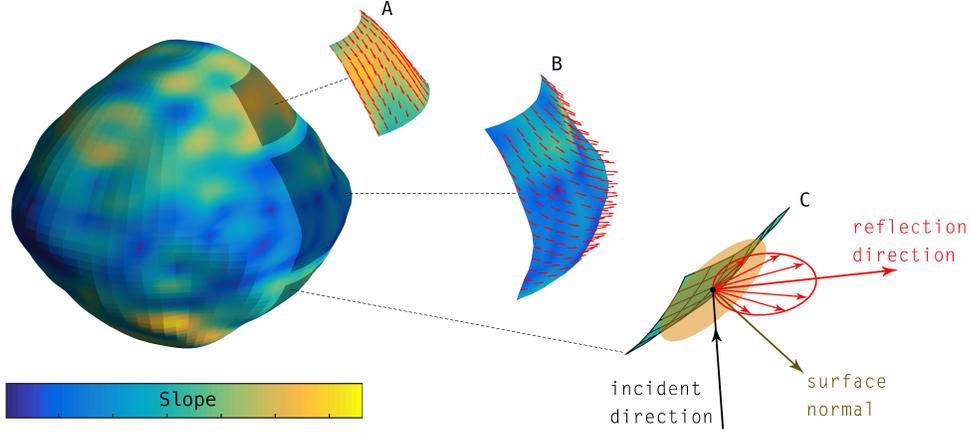}
	\caption{Diagram illustrating the surface mass shedding. The spherical harmonics model (Left; truncated to degree $18$) is created based on the polyhedral shape of $101955$ Bennu, as an example of top-shaped asteroids \citep{pds2019}. The zoom-in view A shows the sliding directions in regions subject to slope failure; B shows the extreme case when the spin rate is exceeding the critical value and the equatorial regions have negative gravity; C shows the stochastic model used to mimic the diffusion of particle rebounding caused by the surface rugosity. \label{f:model}}
\end{figure}

The mass loss caused by rotational instability is highly dependent on local topographic features. We use spherical harmonics to describe the shape of the target asteroid, which provides a good fit to the surface elevation data in a purely analytical form:

\begin{equation}
\label{e:shexpr}
S(\theta,\phi) = \sum_{l=0}^\infty \sum_{m=-l}^l c_l^m Y_l^m(\theta,\phi),
\end{equation}

\noindent in which $c_l^m$ are coefficients of the harmonics and $Y_l^m$ are the bases (e.g., Fig.~\ref{f:model}; see \citet{yu2018} for a complete derivation of the expression). We use a Cartesian coordinate system fixed to the asteroid to locate the parametric shape model, and thus any surface events can be represented in geographic coordinates ($\theta$, $\phi$), \textbf{such as} landslides or a pebble bouncing off \citep{yu2019}.

Assuming the asteroid has homogenous mass distribution, the gravitational acceleration is derived using the polyhedral shape model. Combining the surface gravity, the centrifugal force and the local terrain represented by Eq.~(\ref{e:shexpr}), we derive the steepest descent direction of the effective potential $\bm{l}$:

\begin{equation}
\label{e:slopevec}
\bm{l} = -\nabla V_s + (\nabla V_s \cdot \bm{\hat{n}}) \bm{\hat{n}},
\end{equation}

\noindent in which $V_s$ defines the surface geopotential and $\bm{\hat{n}}$ indicates the unit normal vector of the surface (pointing outwards). The slope is then defined to be the angle between the local geopotential force and the normal direction $<-\nabla V_s, -\bm{\hat{n}}>$, which indicates where landslides mostly likely happen and is used to estimate the local stability of the surface. Considering loose cohesionless particles resting on the surface, and a mean friction coefficient $\mu$, the slope failure is controlled by the condition

\begin{equation}
\label{e:sfcond}
<-\nabla V_s, -\bm{\hat{n}}> = \textup{arctan} \mu.
\end{equation}

\noindent Regions where the slope angles close to or exceeds the friction angle facilitate a local landslide (Fig.~\ref{f:model}A). Debris particles enrolled in landsliding will be accelerated and the mass shedding happens if their peak speeds reach the separation limits of the surface (which depends sensitively on the local terrain; see \citet{yu2018}). In our model, Eqs.~(\ref{e:shexpr})-(\ref{e:sfcond}) are used for analysing the shedding conditions based on an assumption, i.e., non-catastrophic landslides occur sporadically while the asteroid gains relatively high spin rate. This understanding is confirmed by observations of the activity in asteroid Gault \citep{kleyna2019}. These surface events each lasts for a short period compared to the YORP spin-up timescale, e.g., hours to days in millions of years, but the cumulative effects may reshape the system with considerable changes. From a short-term viewpoint, we also assume these day-month-long processes affect mainly the loose cohesionless regolith grains, which cover the subsurface layer with cohesive strength, and do not cause global reshaping effects to the asteroid. The existence of this two-layer structure is supported by recently executed impact experiment on $162173$ Ryugu \citep{arakawa2020}.

Equations~(\ref{e:shexpr})-(\ref{e:sfcond}) are used to monitor the environmental change with respect to the spin rate, i.e., as the asteroid being spun up, the regions at high risk of losing mass spread over the surface. For top-shaped asteroids, such regions spread from the equator to higher latitudes (see \citet{yu2018} for a detailed discussion). An extreme case may occur as the spin rate increases, in which vast regions of the surface correspond to slope angles greater than $90^\circ$ ($\nabla V_s \cdot \bm{\hat{n}} < 0$), so any cohesionless dust/debris in these regions would be levitated due to outward net accelerations (Fig.~\ref{f:model}B). There are actual examples for this case: the observation-based model of the primary of binary asteroid Didymos admits a banded equatorial ridge that is unattachable for materials without tensile strength \citep{yu2019}; and the net accelerations on the equator of $1999$KW$4$ primary are confirmed to be nearly zero \citep{sch2010}.

For modelling the debris/dust shedding field, we use direct Monte Carlo simulations over the unstable regions derived from preceding surface analysis. Given a certain observation time, and a total mass loss speed over the unstable regions, we chose a generation model to determine the launching times, positions and velocities for particles emitted to space. The choice of generation model depends on specific shedding scenario, for example, the emission events are biased by the differences in daylight amount if the thermal stress fracturing acts as a key factor. For the scenario dominated by rotational instability, the emission events are generated based purely on the shedding condition analysis. A certain fraction of the particles will fall back onto the asteroid, and interact with the surface until they are relaunched or accrete on the surface. In this shedding field modelling, debris particles hitting the asteroid surface are reflected diffusely, following a quasi-specular reflection regime (Fig.~\ref{f:model}C). The specular reflected velocity is firstly obtained using an inelastic collision model:

\begin{equation}
\label{e:reflect}
\bm{v}_n^+ = -c_r \bm{v}_n^- , \\
\bm{v}_t^+ = \bm{v}_t^- - \mu \left (1 + c_r \right ) \left \| \bm{v}_n^- \right \| \frac{\bm{v}_t^-}{\left \| \bm{v}_t^- \right \| }.
\end{equation}

\noindent in which $c_r$ is the restitution coefficient. Then a quasi-specular reflection is implemented using the Cercignani-Lampis-Lord (CLL) method \citep{lord1991}. The diffusion is controlled by the roughness of the surface and is adjusted by the energy accommodation coefficient in CLL method. In regions where the shedding condition is false, the particles are allowed to accrete if the hitting speed is below a limit.

\subsection{The discretization of shedding mass} \label{sec:compmass}

For emission induced by rotation, we model the shedding field of the surface with a Monte Carlo resampling methodology. In this method, debris particles are generated following a sampling scheme and assigned with proper launching positions and velocities. The positions are chosen randomly over across the surface area identified by the shedding conditions stated above. And the shedding velocities are assumed to be near-zero \citep{kleyna2019}. For a shedding event that lasts for a couple of days, the average speed of losing mass is controlled by

\begin{equation}
\label{e:sdspeed}
\Delta m = \alpha \Delta t.
\end{equation}

Each unit mass  $\textup{d} m$ should be discretized into the same quantity of particles over the size range of regolith material. Assume the debris particles follow a nominal power law distribution of slope $q$,

\begin{equation}
N_{>d} = N_r d^q,
\label{e:powerlaw}
\end{equation}

\noindent where $d$ is the particle size ($d_l<d<d_u$) and $N_r$ the reference value of number (unit: $1/\textup{m}^q$). The differential form of Eq.~(\ref{e:powerlaw}) yields a number density function in terms of $d$

\begin{equation}
n(d) = -N_r q d^{q-1}.
\label{e:powerlawdiff}
\end{equation}

Define $\rho$ to be the solid density of regolith material, the mass density function in terms of particle size $d$ is

\begin{equation}
\sigma(d) = -\frac{\pi}{6} \rho N_r q d^{q+2}.
\label{e:massden}
\end{equation}

Then the total mass over the size range is determined by integral

\begin{equation}
\int_{d_l}^{d_u} \sigma(d) \textup{d} d = -\frac{\pi}{6} \frac{q}{q+3} \rho N_r (d_u^{q+3} - d_l^{q+3}).
\label{e:totmass}
\end{equation}

A combination of Eq.~(\ref{e:sdspeed}) and (\ref{e:totmass}) give the value of $N_r$:

\begin{equation}
N_r = -\frac{6 (q+3)}{\pi q \rho} \frac{\Delta m}{d_u^{q+3} - d_l^{q+3}},
\label{e:Nrexpress}
\end{equation}

Equation~(\ref{e:Nrexpress}) is used to discretize the mass loss during a differential time $\textup{d} t$. Note $N_r \propto \textup{d} m$, we define a scaled number density $\hat{n}(d) = n(d) / \textup{d} m$. As for the numeric strategy, we assign a series of nodal size values $d_i$ that covers the full size range. The choice of nodal values is guaranteed to capture the various dynamics of shedding debris of different sizes. Then in the Monte Carlo simulation, sampled particles are assigned for each $d_i$, and the sample ratio is defined to be

\begin{equation}
\beta(d_i) = \frac{\hat{n}(d_i)}{n_s},
\label{e:samrt}
\end{equation}

\noindent in which $n_s$ is the sample size that should be assigned with statistically meaningful large value to determine the particles' behaviour for a specific particle size. This numerical scheme identifies the partial surface at high probability of surface instability, and at the same time quantifies the full-scale shedding processes with a distributed size model.

\subsection{Debris cloud propagation} \label{sec:debprop}

The debris materials, after emitted to the circumstance of the asteroid, are exposed to complex mechanical environment and their motion are affected by multiple processes. \citet{yu2017} analysed the perturbations acting on debris particles moving around an asteroid, showing the gravity of asteroid, solar radiation pressure, solar tides and contact forces (inter-particle or between particle and the asteroid) are dominant factors for a month-long evolution. The macroscopic behaviour of the debris cloud essentially depends on the density, or the mean free path of an individual particle. At the start of a shedding event, the debris field is usually rare and its evolution is determined following the ballistic motion of each particle. As the number of particles acumulates and the density becomes large enough, the macroscopic properties of the debris cloud will be governed by the velocity distribution, i.e., inter-particle collisions take over as a major factor to determine the evolution. We adopted a scale-span methodology to analyse this cumulation process. In the first-stage simulation, the shedding particles are propagated using a ballistic model that synthesises all relevant forces based on the previous analysis. The trajectories of particles are calculated in the body-fixed frame of the asteroid using Runge-Kutta integrator, i.e., given the particle's position vector $\bm{r}$, the acceleration vector is defined by

\begin{equation}
\ddot{\bm{r}} =  \bm{a}_g + \bm{a}_e + \bm{a}_c + \bm{a}_t + \bm{a}_s,
\label{e:ballmoto}
\end{equation}

\noindent in which $\bm{a}_g$ is the gravitational acceleration derived from the asteroid's shape model, $\bm{a}_e$ is the convected inertial acceleration, $\bm{a}_c$ is the Coriolis acceleration, $\bm{a}_t$ is the solar tidal acceleration, and $\bm{a}_s$ the solar radiation acceleration (in terms of the particle size). For ash-dust size particles, the magnitude of $\bm{a}_s$ can be far greater than the other items, and the solar occultation matters as it causes an abrupt change to the total acceleration. We consider the intermittent blocking of the solar radiation by using a previously developed algorithm for detecting the intersection of a solar ray \citep{yu2017}.

Monte Carlo simulations are performed to track the motion of particles resampled in Section~\ref{sec:compmass}. The results are used to quantify the statistical features of the debris cloud within the given time span $T$. In particular, the sampled particles of size $d$ determine a spatial distribution, denoted as $\phi (t, \bm{r}; d)$, which defines the number density of $d$-size particles at time $t$ around position $\bm{r}$. Note the distribution function $\phi$ is scaled by a unit mass loss $\textup{d} m$, the cumulative distribution at time $T$ is determined by integral

\begin{equation}
\Phi(\bm{r}; d) = \alpha \int_0^T \beta(d) \phi(t, \bm{r}; d) \textup{d} t.
\label{e:cumdistr}
\end{equation}

Equation~(\ref{e:cumdistr}) gives the number of $d$-size particles placed in a small volume $\textup{d} r^3$ around the position $\bm{r}$ by

\begin{equation}
\Phi(\bm{r}; d)  \textup{d} r^3,
\label{e:numdistr}
\end{equation}

\noindent then combining the numeric results for all nodal sizes $d_i$, we can quantify the change of number density $\Phi$ as a function of size $d$, which allow us to estimate the mass distribution of the debris cloud at time $T$ by

\begin{equation}
\Psi(\bm{r}) = \frac{\pi}{6} \rho \int_{d_l}^{d_u} \Phi(\bm{r}; d) d^3 \textup{d} d.
\label{e:massdist}
\end{equation}

$\Psi(\bm{r})$ defines the mass density function of the debris cloud using an orbital extrapolation. The total mass of the debris cloud at time $T$ is then calculated as a triple integral over the whole domain,

\begin{equation}
M_C = \iiint \Psi(\bm{r}) \textup{d} r^3.
\label{e:cloudmass}
\end{equation}

Equation~(\ref{e:cloudmass}) calculates the total debris mass surrounding the progenitor as a function of time. Given the assumption of a uniform mass shedding in Eq.~(\ref{e:sdspeed}), $M_C$ is monotonous w.r.t time $t$ during the shedding process. We category the dynamical fates of the shedding particles into three types: Recycled (R) -- accrete on the surface of the asteroid;  Leaking (L) -- escape away from the gravitational influence of the asteroid; Cycling (C) -- orbit around the asteroid. $M_C$ comprises the mass of C-type particles, i.e.,

\begin{equation}
M_C \leq \alpha T.
\label{e:mtalt}
\end{equation}

Equations~(\ref{e:cumdistr})-(\ref{e:mtalt}) are used to approximate the heterogenous distributions of the debris cloud, which determines the subsequent evolution of the system. \citet{yu2018, yu2019} examined the evolutionary dependencies of R-, C- and L-types on the physical properties of shedding debris, and showed the irregular gravity field facilitates the clustering of centimeter-size or larger pebbles. We suspect collisions of pebbles on orbit may lead to a collapse phase that is crucial to determine the outcomes of the debris cloud. Thus in the second stage of this methodology, we count in the particle-particle interactions by using the Soft-sphere Discrete Element Method (SSDEM). The total number of particles in the debris cloud is far too large to follow every grain individually in SSDEM. However, we recognize the relatively small particles take the majority of the total number but has only a limited contribution to the total mass, because of the negative slope of the power-law distribution. \citet{yu2018ic} showed the cumulated mass around an asteroid will be highly biased towards large particles over centimeter-level due to the size-sorting effect, indicted by the fast rotation and the solar radiation pressure. Therefore in the SSDEM model we only consider large particles that form the major component of the debris cloud, and track the microscopic dynamical behaviours of the granular media.

The SSDEM simulations are performed using an original code, DEMBody. DEMBody is developed for planetary science and geophysics, and has been employed for reconstructing the evolutionary history of celestial bodies, whose results highly coincide with the observations \citep{cheng2021,cheng2022,huang2023}. The nonlinear contact forces between two contacting spheres are calculated according to the Hertz-Mindlin-Deresiewicz contact theory \citep{somfai2005}, whose components include the normal / tangential forces and the rolling / twisting torques \citep{mehta2007,cheng2018}. Besides the elastic, damping and friction effects in the contact points, the mesoscale cohesive effect is also considered following the granular-bridge theory developed by \citet{sansch2014}, and the cohesion force is approximated by

\begin{equation}
F_c = c \frac{(\beta \bar{d})^2}{4},
\label{e:cohef}
\end{equation}

\noindent in which $\bar{d}$ is the effective particle size, $c$ is the inter-particle cohesive strength and $\beta$ is the shape parameter that represents the statistical deviation from a sphere \citep{zhang2018}, which controls the rotational resistance to relative motions in rolling and twisting directions \citep{jiang2015}. We applied the polyhedron gravity to the granular system to ensure the consistence in mechanical environment. The solar radiation pressure is omitted because it is in a negligible level compared with other forces.

In our SSDEM simulation, we initialize the granular system using the cumulation results exported from the first-stage simulation. By selecting large particles above a threshold value, the system size is reduced to an acceptable scale that is executable on most high performance computing platforms. The motion of the granular system is then propagated using a second-order leapfrog integrator in small time steps. The second-stage simulation is performed to sweep a wide range of time, which is confirmed to be sufficient to enable a complete observation of the transition of the whole system. The choice of stopping time depends on the macroscopic property of the granular system, especially whether the energy supply and dissipation reach a balance to maintain the whole system a steady state.

\section{Results} \label{sec:result}

\subsection{Simulation setup} \label{sec:simset}

Previous studies showed an evolutionary dependence of the shedding debris on the morphology of the progenitor body. However, observations of the fast-rotating targets that have records of shedding events can not offer additional information on shape modelling. To perform a numeric investigation of the post-shedding evolution, we have to design a hypothetical scenario using a target with reliable shape model. In this section, the Near-Earth Asteroid $162173$ Ryugu is employed as an example. It has a typical top-like overall shape and Japan's Hayabusa2 mission has returned high-fidelity information of this target, e.g., the shape, gravity, bulk density, rotational states, and surface regolith composition, etc.

\begin{table}[h!]
\centering
\begin{threeparttable}[b]
\caption{Directly measured dynamical parameters of 162173 Ryugu}
\label{tab:dynpara}
\vspace{0.05in}
\begin{tabular}{lll}
\hline
\hline
Parameter 				& Value 								& Reference			\\
\hline
Mean diameter				& $0.896 \pm 0.004$ km 					& \cite{watanabe2019} 	\\
Bulk density				& $1.19 \pm 0.03$ g/cc 					& \cite{watanabe2019} 	\\
Rotation period				& $7.627 \pm 0.007$ h 					& \cite{abe2018}		\\
Heliocentric pole direction         & ($178^\circ$, $-87^\circ$)				& \cite{benner2020} 		\\
Shape model 				& SFM$20180804$\tnote{1}				& \cite{wata2019sci} 		\\
Cumulative size distribution	& ($-2.65 \pm 0.05$, $>5.0$ m)\tnote{2}		& \cite{michikami2019}	\\
						& ($-2.07 \pm 0.06$, $0.2-9.1$ m)\tnote{3}	& \cite{michikami2019}	\\
						& ($-2.01 \pm 0.06$, $0.1-4.1$ m)			& \cite{michikami2019}			\\
						& ($-1.96 \pm 0.07$, $0.05-3.37$ m)			& \cite{michikami2019}			\\
						& ($-1.98 \pm 0.09$, $0.02-1.83$ m)			& \cite{michikami2019}			\\
						& ($-1.65 \pm 0.05$, $0.02-3.26$ m)			& \cite{michikami2019}			\\						
\hline
\end{tabular}
\begin{tablenotes}
	\item[1] Original SFM model has $3,145,728$ triangular facets, which was remeshed into $5,120$ facets in this study when calculating the gravity of asteroid.
	\item[2] The bracket indicate the power index of the size distribution for boulders larger than $5$ m, obtained from the images of Ryugu's surface.
	\item[3] The power indexs of the size distribution in respective size ranges, obtained by MINERVA-II lander of Hayabusa2.
\end{tablenotes}
\end{threeparttable}
\end{table}


Table \ref{tab:dynpara} presents the parameters of Ryugu measured directly by Hayabusa2 mission. Based on an assumption of homogenous interior mass distribution, the critical rotation period is estimated to be $\sim 3.37$ h, which can cause a local surface failure on the equator. Several observed shedding events can be used to quantify the properties of shedding mass from a fast rotating asteroid.  Table \ref{tab:shedobsev} listed four events associated with the main-belt asteroids $6478$ Gault and P/2013 P5, respectively. Streamers corresponding to dust releasing were recorded by optical observations for these events \citep{hainaut2014,kleyna2019}. Ground-based data show in these events the releasing speeds are negligible and Gault has a spin rate close to the rotational limit (for P5 there is currently rare information about rotation but a critical spin rate cannot be ruled out), which makes the YORP-related instability the most likely explanation to these shedding events.

\begin{table}[h!]
\centering
\caption{Typical shedding activities of main-belt asteroids likely caused by rotational breakup \citep{hainaut2014,kleyna2019}}
\label{tab:shedobsev}
\vspace{0.05in}
\begin{tabular}{llllll}
\hline
\hline
Event \#		& Object			& Time		& Esti. mass loss		& Esti. size constitution 			& Duration \\
\hline
1			& 6478 Gault		& 12/30/2018	& $7 \times 10^9$ kg		& ($-1.70$, $0.03-2$ mm)			& 15 days \\
2			& 6478 Gault		& 12/08/2018	& $4 \times 10^7$ kg		& ($-1.64$, $0.03-2$ mm)			& 5 days \\
3			& P/2013 P5		& 06/14/2013	& $3 \times 10^6$ kg		& ($-1.0$, $1-1000$ $\mu$m)		& 50 days \\
4			& P/2013 P5		& 07/22/2013	& $2.6 \times 10^7$ kg	& ($-1.0$, $1-1000$ $\mu$m)		& 20 days \\					
\hline
\end{tabular}
\end{table}

The mass loss in Table \ref{tab:shedobsev} is quantified using the measured grain size distribution, which is formulated as a power-law fit over the observed size range (the 5th column in Table  \ref{tab:shedobsev}). Assuming the grain density $\rho = 2.848$ g/cc for both objects, the estimated mass loss is calculated following Eqs.~(\ref{e:massden})-(\ref{e:totmass}). Note that the estimated value represents only a small fraction of the actural mass loss, because the majority of the shedding mass is dominated by large particles exceeding the sizes listed in Table 2, which are rarely present in photometric observations of streamer. \cite{yu2019} showed dust/sand-sized particles (below millimeter level) around an $800$-meter-sized asteroid will be dispersed rapidly by solar radiation pressure and cause the formation of the trails; the pebble/cobble-sized particles (above centimeter level) will be bound to the asteroid for a longer term and thus make little photometric contribution to the streamer. The latter is small in number but take the majority of the shedding mass.

Differing from the mass loss from an impact process, the shedding process depends significantly on the progenitor's size. In the scenario of Ryugu, a rough approximation of the shedding mass can be obtained from these references by defining a scaled index as

\begin{equation}
\beta_m = \frac{M_{<d_l,d_u>}}{M_A},
\label{e:scalemass}
\end{equation}

\noindent in which $M_{<d_l,d_u>}$ is the shedding mass of particles within size range $<d_l,d_u>$, and $M_A$ is the total mass of the asteroid. The events in Table \ref{tab:shedobsev} determine $\beta_m$ ranging from $4.0 \times 10^{-7}$ to $1.4 \times 10^{-4}$, corresponding to the measured grain size from micrometer to millimeter. Another factor controlling the mass leaking magnitude is the ratio of the actual shedding mass to the value of measured size range, denoted by $\kappa$:

\begin{equation}
\kappa = \frac{M_\textup{loss}}{M_{<d_l,d_u>}}.
\label{e:mratio}
\end{equation}

The ratio $\kappa$ is obviously a positive number over $1.0$ but has a high uncertainty because our knowledge is very limited about the mass of little photometric contribution. This is a purely assumed number in this study, but there is still constraint from the total mass of the asteroid. We consider a shedding event that doesn't cause a global reshaping effect, which should not lead to a significant mass loss compared to the asteroid, i.e., $\kappa \beta_m << 1.0$.

In the first-stage simulation, we consider a scale number $\beta_m$ ranging from $10^{-7}$  to $10^{-4}$, and the duration of the shedding process is assumed to be $20$ days, due to the reference values in Table~\ref{tab:shedobsev}. The rotational period is set to $3.2$ hr, which is slightly above the critical spin limit. In this scenario, the polar areas are mostly stable; an unstable strip area emerges around the equator, where the centrifugal acceleration exceeds the surface gravity. So once a cohesive failure happens the loose material should be rapidly cleared from this region, leaving unweathered subsurface exposed. Tracer particles are sampled randomly across the unstable area following the scheme described in Section~\ref{sec:surfdyn} and Section~\ref{sec:compmass}. The launching velocities of tracer particles are assumed to be zero. Then the propagation of debris cloud released from a shedding event is simulated using the methodology of Section~\ref{sec:debprop}.

\subsection{Characterization of the debris cloud} \label{sec:chrctcld} %

We first describe the general features of the cumulated debris cloud during the shedding process. Previous studies show the fates of shedding particles exposed in the near-field mechanical environment exhibit strong dependence on the particle size. The solar radiation pressure, interfered by the asteroid occultation, produce complex influence on the ballistic motion of the debris particles. \citet{yu2019} examined the expansion of shedding particles from fly ash to cobble size, and showed the clearance of particles below 100 $\mu$m will be in 2 days for asteroid Didymos A ($\sim$ 800 m in size, close to Ryugu). Therefore we can expect a strong size segregation during the shedding process. We consider a size range $100$ $\mu$m-$10$ m based on the analysis and observations, and a power index $q$ between $-2.65$ and $-1.65$ \citep{michikami2019}. It is an arbitrary choice to use a constant $q$ for the full size range, nevertheless, there is no validated model that spans all data sets so far \citep{phyden2020}. We thus adopt a simple power law distribution with variant power index, which allows us to understand the post-shedding evolution in terms of the size composition of the regolith material.

\begin{figure}[h!]
\centering
\includegraphics[width=0.98\textwidth] {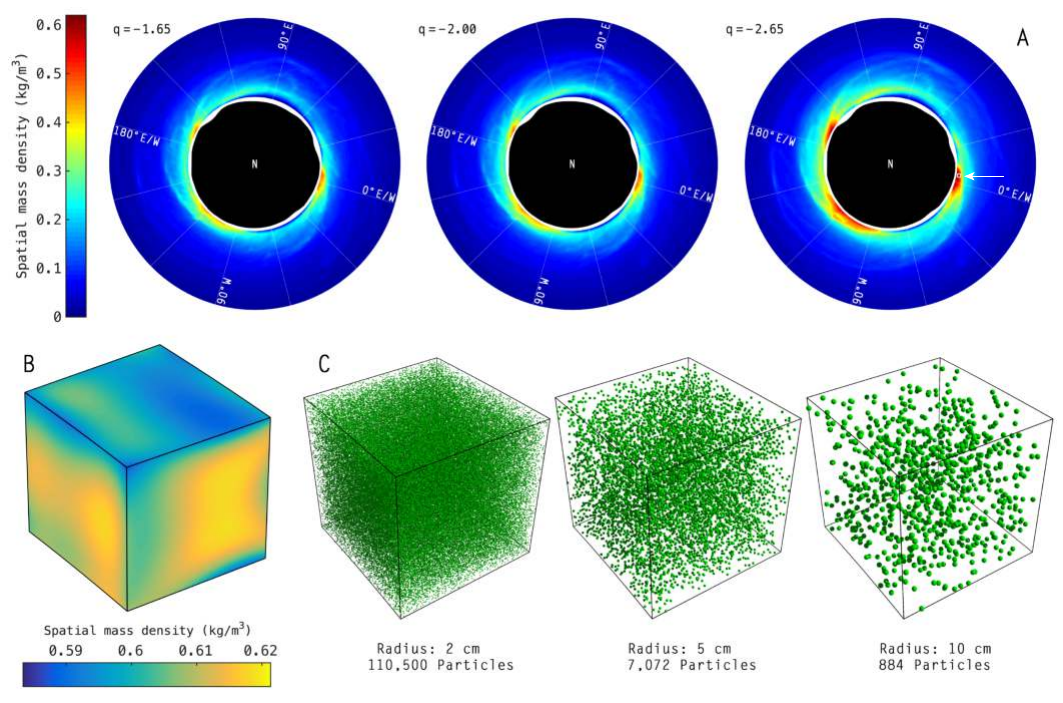}
\caption{A. The mass distribution of the cumulated debris cloud at the end of the 20-day-long shedding process. The colour spectrum indicates the spatial mass density of the equatorial plane plotted in the body-fixed frame (Left: $q=-1.65$; Middle: $q=-2.0$; Right: $q=-2.65$). We adopted the geographic coordinates defined by \cite{hira2019}. B. A zoom-in view of the cube area with peak mass density, which is marked in the rightmost cloud diagram of panel A using a white solid wireframe pointed by a white arrow. C. Discretization results of the zoomed area in panel B (with identical total mass), i.e., three uniform particles sizes, $2$ cm, $5$ cm and $10$ cm, are considered for examples ($4$ times of the original sizes are drawn for visual enhancement). An animation of the evolution of cumulative spatial mass density for power index $q = -2.0$ is available. The animation lasts for 28 s and describes the 20-day shedding and propagating process in the body-fixed frame of the asteroid, whose last frame corresponds to the middle plot of panel A. \label{f:mass2q}}
\end{figure}

Figure~\ref{f:mass2q}A illustrates the spatial mass density at the end of the shedding process for three power indexes $q=-2.65$, $-2.0$ and $-1.65$, which covers the measured values as listed in Table \ref{tab:dynpara}. Equation~(\ref{e:massdist}) indicates the mass density is proportional to the mass shedding ratio $\beta_m$, and as representative Fig.~\ref{f:mass2q}A calcultes the results for a ratio $\beta_m = 7.2\times10^{-5}$. The profile section of the debris clouds are projected to the polar grid of the equatorial plane, where the mass densities are colored with a uniform spectrum. Although the evolutionary path of a single debris particle depends strongly on the initial conditions and particle size, our simulation shows the cumulative behavior of the shedding mass exhibit almost identical temporal-spatial patterns for different size compositions. We consider 20 days as a reasonable interval to scatter the shedding particles, whose trajectories are perturbed by spin-orbit resonance, the solar radiation pressure and collisions with the asteroid surface. The cumulative distributions of the shedding mass for all three considered power indexes show invariant tendencies: first, the cumulative mass is concentrated in the vicinity of the asteroid and gets sparse rapidly as the distance increases from the surface; second, the cumulated debris cloud distributes unevenly around the asteroid at the same radial distance, e.g., for Ryugu, three high-density regions emerge close to the surface, and such regions remain their body-fixed locations during the whole post-shedding process; third, the concentrated regions exhibit structural stability that depends barely on the size composition $q$ and shows no obvious correlation to the local terrain of the asteroid. The reader is referred to the online version of this article for a complete movie of the debris cloud evolution (a 20-day evolution of the cumulative mass is visualized in the body-fixed frame of the asteroid for power index $q = -2.0$ by the animation of Fig.~\ref{f:mass2q} in the online article).

We took a detailed look at the concentrated region of maximum mass density (the boxed area in the right panel of Fig.~\ref{f:mass2q}A; $q=-2.65$). Figure~\ref{f:mass2q}B is a zoomed view of the marked volume; the sections of the cubic area are colored in a narrow-scope spectrum, showing the fluctuation of the spatial mass density, which peaks at $0.62$ kg/m$^3$. For the interests in checking the cumulative effects of the shedding particles, Fig.~\ref{f:mass2q}C listed the discretization results of the boxed area using uniform particle sizes, i.e., $2$ cm in radius for the left cube, $5$ cm for the middle cube, and $10$ cm for the right cube. Figure~\ref{f:mass2q} gives an intuitive sense on the cumulative effects of shedding particles that cluster in the vicinity of the progenitor, i.e., a reasonable shedding magnitude $\beta_m = 7.2\times10^{-5}$ corresponds to a peak mass density $\sim 0.6$ kg/m$^3$ given a solid density $2.848$ g/cc. 

\begin{figure}[h!]
\centering
\includegraphics[width=0.98\textwidth] {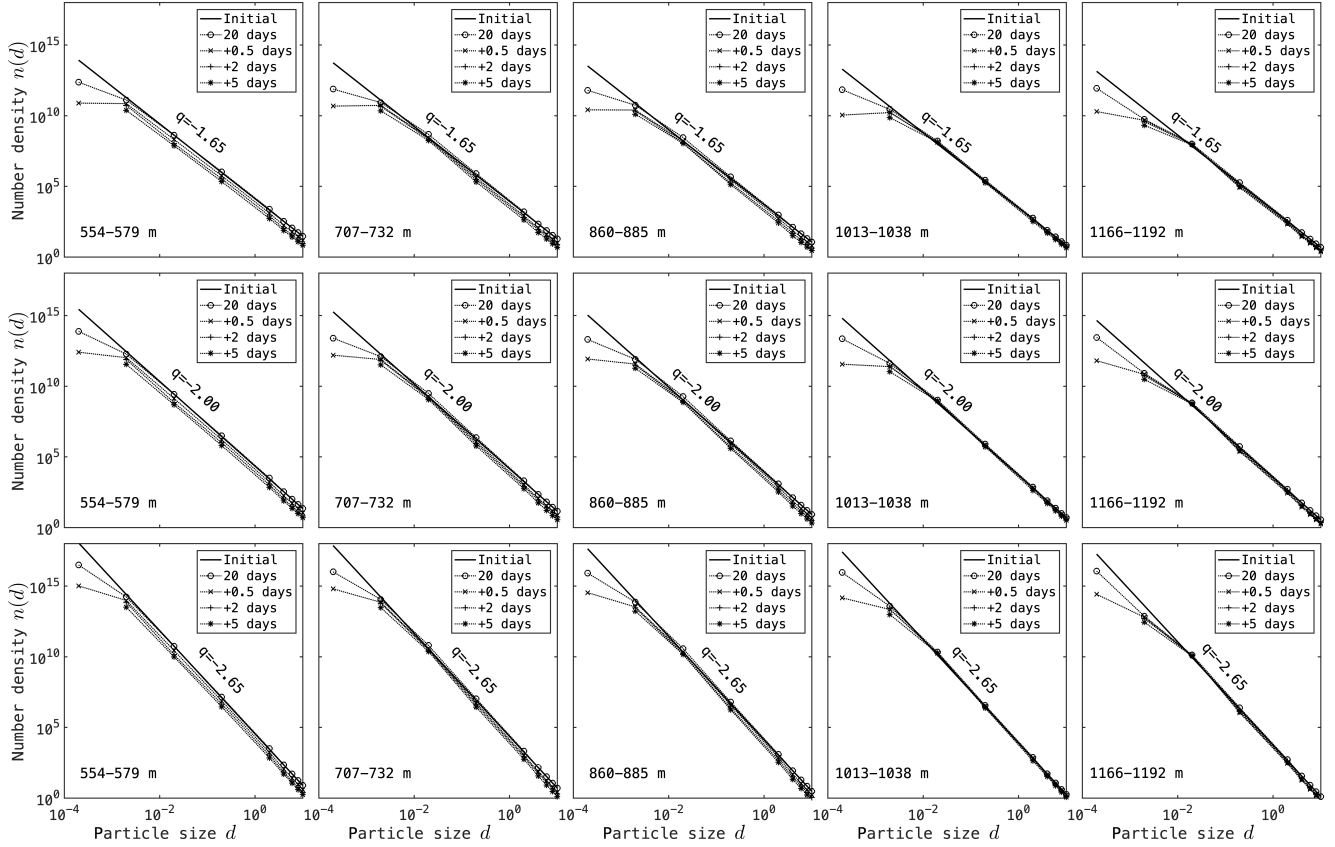}
\caption{The size segregation effect in terms of the initial size distribution and the radial distance from the asteroid. Each panel calculates the result of a specific distance range for a given power index, presented in a log-log plot. The solid lines indicate the original size distribution of the assumed regolith material, and the dotted lines with markers indicate the size distributions at different time points after the shedding process. \label{f:sizecomp}}
\end{figure}

A comparison among the results of three power indexes show the total cumulative mass remains a steady level. Figure~\ref{f:sizecomp} illustrates the size segregation effect in terms of the initial power index and the radial distance. Each panel in Fig.~\ref{f:sizecomp} shows a log-log size distribution for a specific range of the radial distance, changing as a function of time. The fitted size distributions at $5$ time points are checked in the figure, i. the initial indexes our numerical experiments, i.e., $q=-2.65$, $-2.0$ or $=1.65$; ii. at the end of the shedding process; iii. $12$ hours after the shedding process; iv. $2$ days after the shedding process; v. $5$ days after the shedding process. Some simple facts are noted: First, the size segregation is quite weak for particles over the centimeter level, thus the fitted power indexes are barely shifted for pebble-cobble sizes. Second, strong size sorting effects are observed for particle below the milimeter level, which shows little dependence on the radial distance. $10^{-4}$-meter-level particles are cleaned from the vicinity of the asteroid in $2$ days, and the milimeter-level particles show continuous dropping as time after the shedding process. The results suggest for a fast-rotating asteroid below $1$ km, the fine grains of the shedding mass contribute little to the cumulation of a steady debris cloud around it, therefore the formation of debris aggregates will be dominated by collisions between large pieces of debris.

\begin{figure}[h!]
\centering
\includegraphics[width=0.5\textwidth] {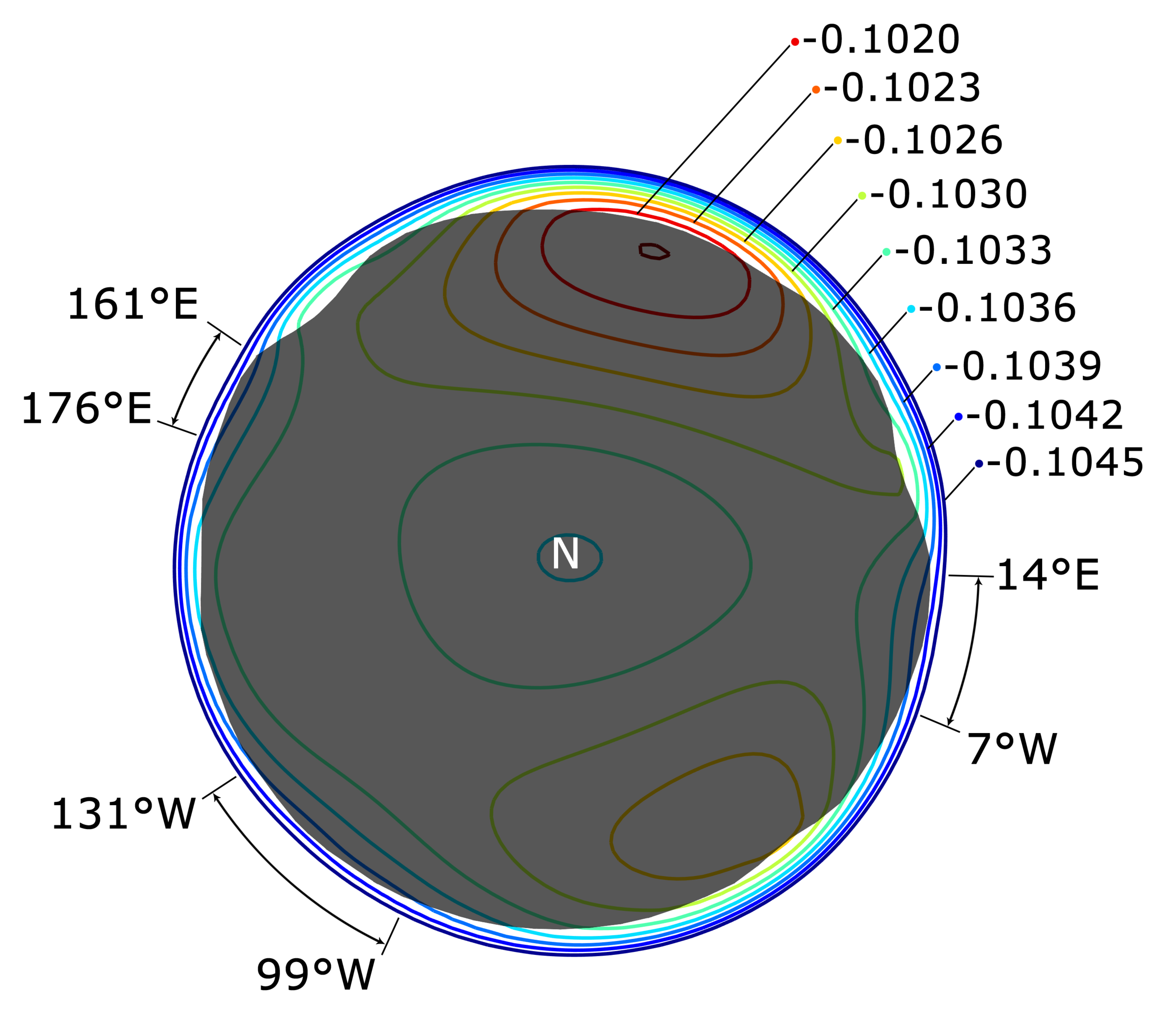}
\caption{Contour lines of the geopotential near the surface of the asteroid projected to the equatorial plane. The countour lines are marked in different colours indicating their geopotential values (unit: Joule per kg). The shadowed area indicates the polar view shape of Ryugu. \label{f:surfpotential}}
\end{figure}

Uneven distribution of the cumulative debris cloud is noticed in the vicinity of the asteroid. At the end of the shedding process, a large majority of the debris mass is compressed within a range from $1.0$ to $1.2$ times of the asteroid radius. And a steady mass structure emerges in the cumulated cloud, which shows little dependence on the size composition of the shedding mass (Fig.~\ref{f:mass2q}A). Three regions of evident mass concentration locate in longitudes $131^\circ$ W - $99^\circ$ W,  $7^\circ$ W - $14^\circ$ E and $161^\circ$ E - $176^\circ$ E; and regions of relative dilute mass locate in $60^\circ$ W - $30^\circ$ W and around $90^\circ$ E. Such a steady mass structure is correlated with the complex dynamical environment close to the asteroid surface, which is individual and different for different progenitors. From a local perspective, the mass distribution is determined by the flux of debris particles. The asteroid model employed in our example simulations has unattachable equatorial ridge due to the fast rotational speed, thus the shed debris particles move over the surface in a particular way called ``saltation'' \citep{harris2009}, meaning that the migration of particles relies on complicated coupling effects with the asteroid surface which will affect the influx and outflux significantly. We calculated the geopotential (defined to be the sum of the gravitational potential and the centrifugal potential) near the surface of the asteroid, in order to find the topological structure that may be responsible for the observed general trend of the migrating particles. Figure~\ref{f:surfpotential} illustrates the contour lines of the geopotential on the equatorial plane near the asteroid, plotted for the critical rotational period $3.2$ hr. Curves in different colours are denoted by their values of geopotential, showing a fluctuation on the asteroid surface, which is inconsistent with the fluctuation in elevation. Remarkably, we found the surface valley areas of the geopotential are in perfect agreement with the mass concentration as stated above, and the peak areas are in agreement with the mass vacuum regions.

The numeric results confirmed our speculation that the shed particles tend to migrate from the high geopotential to low geopotential via saltatory processes on and across the asteroid surface, and such processes determine the basic trend of the cumulative debris cloud near a fast-rotating asteroid. Towards an understanding of this trend, we consider a debris particle moving across the asteroid surface through a saltatory route. Since the collisions with the surface are imperfectly elastic, every contact with the surface will cause dropping of the relative kinetic energy, i.e., the first term of Eq.~(\ref{e:geopdef}). The geopotential is denoted by $V$ and the Jacobi integral of the particle's orbital motion can be written as

\begin{equation}
J = \frac{1}{2} \dot{\bm{r}} \cdot \dot{\bm{r}} + V(\bm{r}),
\label{e:geopdef}
\end{equation}

\noindent where $\dot{\bm{r}}$ is the relative velocity of the particle w.r.t. the body-fixed frame. The integral constant $J$ maintains invariant for a non-collisional ballistic motion due to \citet{sdj1996}, and each collision with the surface cause a reduce in the integral. Thus the saltation route leads to a monotonous decreasing $J$ towards some limit states, which are the local minimum regions of the geopotential as illustrated in Fig.~\ref{f:surfpotential}. We note such a process does not depend on the initial position where the shedding occurs, so all shedding particles will be transferred to the local minimum geopotentials, and later enter long-term survival orbits that are not intersected with surface of the asteroid. It explains the steady mass structure of the cumulative debris cloud, in which the mass concentration is a result of dense influx by the saltatory transportation.

We note the saltatory transportation is prevalent for asteroids approaching their critical spin limits, and the only condition is uneven surface topography that shaped a fluctuation in the surface geopotential. As a common mechanism we believe it plays a role in promoting the aggregation of shedding particles and hence facilitates the collisional growth of debris near a fast rotating asteroids.

\subsection{Handing-off to DEM simulations} \label{sec:handoff} %
As the particle flow grows denser, inter-particle collisions will increase, which dominate the subsequent evolution of the debris cloud. Thus we developed a hand-off method from ballistic model to discrete element method for tracking the evolution of a dense flow. In this section, we aim to justify the rationale for switching models and provide a comprehensive overview of the simulation strategy for the entire process. First, we simulated the debris propagation of a single mass shedding event through employing the ballistic model and DEM separately to prove that the ballistic model has sufficient accuracy in capturing the global structure of the shedding and propagating phase, and its low computational cost enables us to check a wide particle size range and a huge particle number over the evolution of several months. Then we give a coherent picture illustrating the simulation strategy of the whole evolutionary process of the debris cloud, from shedding and propagating to clustering.

The ballistic model calculated the propagation of the shedding particles in terms of Eq.~(\ref{e:ballmoto}), and also considered the collisions with asteroid surface using the reflection method of Eq.~(\ref{e:reflect}). Unlike the ballistic model, the DEM simulation incorporates the inter-particle gravity, contact, collision, and cohesion. Considering the computational consumption of the DEM simulations, we simulated the detachment and dispersion of over 80,000 particles (each 10 cm in size) from the unstable regions over a period of 3 days in a single mass shedding event. We found that the propagating paths and the temporal-spatial distribution of the shed particles obtained from the two computational models are similar, which indicates that the effects of colliding and clustering are minimal while the geopotential is predominant in the shedding and propagating phase.

\begin{figure}[h!]
	\centering
	\includegraphics[width=1.0\textwidth] {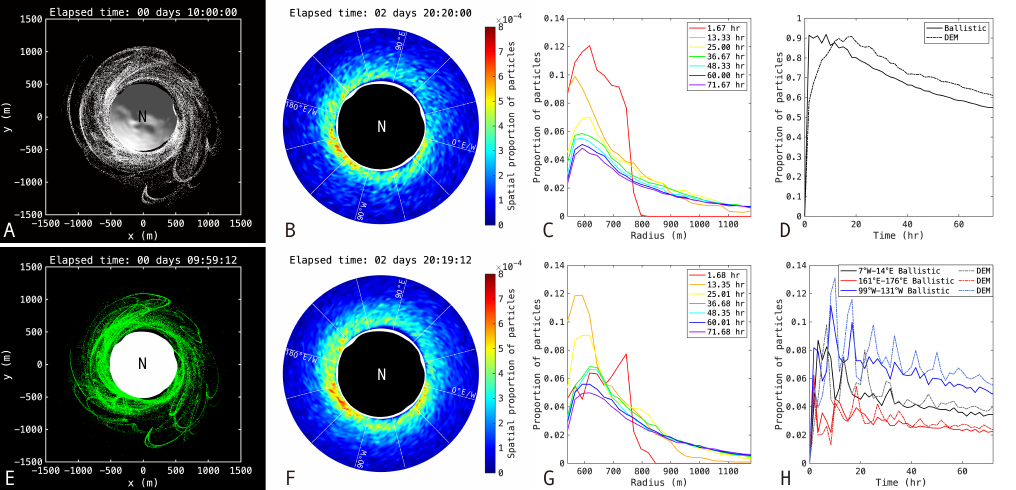}
	\caption{Comparison of the ballistic model and the discrete element method in simulating the dispersion process of shed debris. Panels A, B and C are from the ballistic simulation, and panels E, F, and G are from the DEM simulation. A and E show the propagating patterns of the shed particles after 10 hours of the shedding event. B and F exhibit the proportion distribution of particles within grids defined by longitudinal and radial nodes around 2 days, 20 hours and 20 minutes later. The cumulative particle distribution in the z-direction is merged and displayed in the proportion distribution map in the equatorial plane. C and G present the distribution of particle proportions along the radial distance at different time points. D. The proportion of particles entering the annular area between the radial distances of 528 meters and 1,200 meters, calculated by the two models. The annular area is shown as the cloud map in the panels B and F. H. The proportion of particles within three low-geopotential regions changes over time, with solid lines representing the results from the ballistic simulation and dash-dot lines representing the DEM simulation.\label{f:comparison}}
\end{figure}

Due to the different calculation time steps used in the ballistic simulation and the DEM simulation, we compared the results of the two models at moments that were as close as possible to each other. Figure~\ref{f:comparison} shows the measurements of propagating characteristics from the two computional models. Figure~\ref{f:comparison}A and \ref{f:comparison}E present the dispersion state of particles $\sim10$ hrs after the onset of particle shedding from the top view, whose data are from the ballistic simulation and DEM simulation respectively. It is evident that the spiral structures in both figures are similar, and the global mass distributions are consistent. 
We compared the spatial distribution of particles throughout the 3-day simulations. The proportion of particles within grids that are divided by longitudinal and radial nodes is measured as shown in Fig.~\ref{f:comparison}B and \ref{f:comparison}F. At the end stage of the simulations, the computational results from both models exhibit the same regions of particle concentration and particle voids, generally consistent with the dense and dilute regions of the 20-day shedding evolution of multi-disperse particles in Section~\ref{sec:chrctcld}. 
We also analyzed the dispersion of particles within the annular region spanning radial distances from 528 m to 1200 m. The region is as depicted in the cloud maps in Fig.~\ref{f:comparison}B and~\ref{f:comparison}F. As shown in Fig.~\ref{f:comparison}C and~\ref{f:comparison}G, there are differences in the dispersion speeds of particles on the first day of the two models' simulations: In the ballistic simulation (Fig.~\ref{f:comparison}C), the majority of particles quickly enter the annular area within 800 m (as shown by the distribution curve at 1.67 hr), then gradually disperse outward, as indicated at 13.33 hr and 25.00 hr. Subsequently, the total amount of particles remaining within this area slowly decreases, but the concentration within 800 m remains high. At 1.68 hr, the DEM simulation (Fig.~\ref{f:comparison}G) does not show as many particles falling into the 528$\sim$800 m range as in the ballistic simulation, and the remarkable concentration in the vicinity of the asteroid appears until 13.35 hr. After 36 hrs, the radial distribution of particles in both models shows similarities.
In Fig.~\ref{f:comparison}D, the temporal evolution of the total particle count within the 528$\sim$1200 m annular area also displays the previously mentioned differences and similarities: In the ballistic simulation, the majority of shed particles quickly enter this area; in contrast, the increase in particle amount entering this area in the DEM simulation is slower. This discrepancy stems from the initial settings of the simulations: in the ballistic simulation, all particles in the unstable area are released at the same moment, whereas in the DEM simulation, particles adhered to the unstable surface are released gradually. Excluding the differences caused by the initial settings, the decrease in the proportion of particles within the annular area during the latter half of simulated period shows a similar slope in both models. 
The evolution of particle proportions within three mass concentration zones (low geopotential zones) mentioned in Section~\ref{sec:chrctcld} is scrutinized (Fig.~\ref{f:comparison}H). The DEM simulation reveals more pronounced fluctuations in particle proportions in the first half of the period, attributed to the inter-particle gravity, collision dissipation and cohesive effect. These forces or effects promote the collective behavior as particles traverse the observed regions. This intense fluctuation gradually diminishes later on, and the particle proportions in these three zones begin to stabilize. Despite these fluctuations, the overall trends in particle proportions across these areas are consistent when analyzed with both computational models. 
The comparative analyses across various aspects conclusively show that during the particle shedding and propagating phase, the results from the ballistic and DEM models are statistically consistent. Interactions such as contact, collision, and cohesion between particles are not dominant during this phase. Instead, the primary factor influencing particle dispersion paths and distribution patterns is the asteroid's irregular gravitational field, which is the main common element in both models. Consequently, the ballistic model, which requires less computational effort, provides sufficient accuracy for simulations of the shedding and propagating phase.

\begin{figure}[h!]
	\centering
	\includegraphics[width=0.5\textwidth] {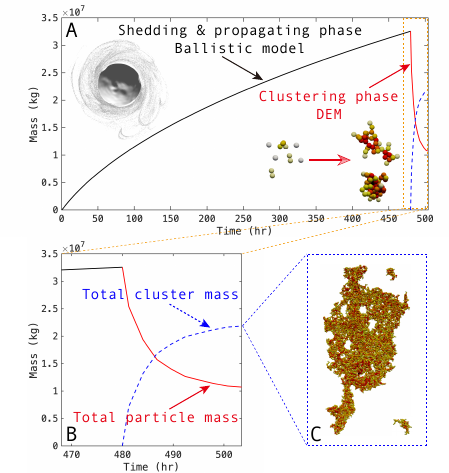}
	\caption{A coherent picture illustrates the complete simulation strategy from the shedding events to the steady state of clustering phase. The black solid line represents the total mass of particles that have detached from the surface and dispersed in the orbit over 20 days, calculated by the ballistic model. The red solid line indicates the total mass of free particles in orbit that have not formed clusters, with the clustering phase simulated through DEM and lasting 1 day. The blue dashed line shows the total mass of all clusters formed by particle cohesion, as illustrated in panel C. Panel B provides a magnified view of the mass evolution at the end of the first phase and during the second phase.\label{f:coherentpic}}
\end{figure}

When the spatial density of particles in the debris cloud reaches a critical threshold, increasing the frequency of particle collisions, the collisional growth of clusters becomes non-negligible. At this juncture, we transition to using the discrete element method to capture the growth of clusters within the debris cloud, which are potential precursors to asteroid moons. Figure~\ref{f:coherentpic} illustrates the comprehensive simulation strategy employed in this study. The first phase involves mass shedding over 20 days, calculated by the ballistic model to simulate the propagation and dispersion of shedding particles. The black solid line in Fig.~\ref{f:coherentpic} represents the accumulation of total particle mass in orbit during this period. The mass shedding is designed to continue for 20 days, in accordance with observed facts as listed in Table~\ref{tab:shedobsev}, which will result in a continuous increase in the spatial mass density of the debris cloud. Collisional growth among orbiting particles gradually becomes prevalent, which requires discrete element method to capture the clustering process. So the simulation progresses into the second phase, handing off to the discrete element method. The red solid line shows the continuous decrease in the mass of free particles in orbit, with the mass reduction transitioning into cluster mass, depicted by the blue dashed line. After 24 hrs, the total mass of the clusters nearly reaches a stable state. The details of the particle accretion into aggregates during the clustering phase will be discussed in Section~\ref{sec:accana}. 
This study presents a feasible and rational two-stage simulation approach for investigating the evolution of shed debris from unstable asteroids, which depicts a possible path toward the formation of asteroid moonlets. This section justifies the availability of the transition from a 20-day ballistic simulation to a 1-day DEM simulation, although the optimal timing for switching between the two models requires further investigation through systematic simulations, which will be one of the downstream tasks of this work.

\subsection{Accretion analysis} \label{sec:accana} 

Three separate DEM simulations were performed following the 20-day-long shedding process using the hand-off model. The mass distribution of the debris cloud as demonstrated in Section \ref{sec:chrctcld} is first discretized into a non-contact particle cloud that corresponds to the same mass structure. Three particle sizes $2$ cm, $5$ cm and $10$ cm are employed in the simulations, which are chosen because $2$ cm to $10$ cm grains are directly measured on the regolith of multiple subkilometer asteroids \citep{lauretta2022,tachibana2022,walsh2022}. For comparison we use the preliminary results of $q=-2.65$  for all three simulations. The converted number of $10$ cm particles (with identical mass) is $2,728,342$, which is massive but acceptable for the parallel code DEMBody; for $5$ cm, the number is $21,826,736$, and for $2$ cm, is $341,042,750$, which are far beyond the limit of the code processing. We have to shrink the sample size by deleting regions where the particle cloud is dilute and only considering the densest regions (see Fig.~\ref{f:mass2q}A), thus the simulated samples can be limited around the million size.

Table~\ref{t:dempara} lists the particle contact parameters used in our DEM code, which are chosen to mimic the contact behaviors of regolith particles launched from subkilometer asteroids. We adopted a softened value for the Young modulus $Y$. It is a technique frequently employed in DEM simulations, which allows for an acceptable time step on the premise of the largest inter-particle overlap $\delta_\mathrm{max}$ below $1\%$ of the particle radius \citep{zhang2018}. The friction coefficient $\mu_\mathrm{s}$ and $\mu_\mathrm{d}$, the restitution coefficient $\epsilon$, and the shape parameter $\beta$ are determined according to the properties of sand particles with medium hardness and repose angle of $\sim30^{\circ}$ \citep{jiang2015}. The inter-particle cohesive strength $c$ is applied to make the macroscopic granular clusters possess tensile strength. The initial positions and velocities of the particles are obtained from the interpolation of the ballistic model. Special interests are paid to the collisional growth of the debris particles, aiming at a detailed look into the early process of the small companions' formation. We track the evolution of the debris cloud for a simulated time of $24$ hrs and locate all aggregates (defined as the union sets of debris particles that are connected via contact forces) changing as functions of time.

\begin{table}[h!]
\centering
\caption{Particle contact parameters in the SSDEM code DEMBody, which are chosen to represent the mechanical properties of the regolith particles on subkilometer sized asteroids}
\label{t:dempara}
\begin{tabular}{@{}lcl@{}}
\hline
Parameter    					& Symbol 			& Value\\
\hline
Young modulus			& $Y$ 			& $5\times10^3$ Pa\\
Poisson's ratio			& $\nu $ 			& $0.3$\\
Static friction coefficient  		& $\mu_\mathrm{s}$ 		& $0.6$ \\
Dynamic friction coefficient  		& $\mu_\mathrm{d}$ 		& $0.6$ \\
Restitution coefficient		& $\epsilon$ 		& $0.5$ \\
Shape parameter			& $\beta$ 		& $0.6$ \\
Cohesive strength			& $c$ 		& $2\times10^2$ Pa\\
\hline
\end{tabular}
\end{table}

\begin{figure}[h!]
	\centering
	\includegraphics[width=0.75\textwidth] {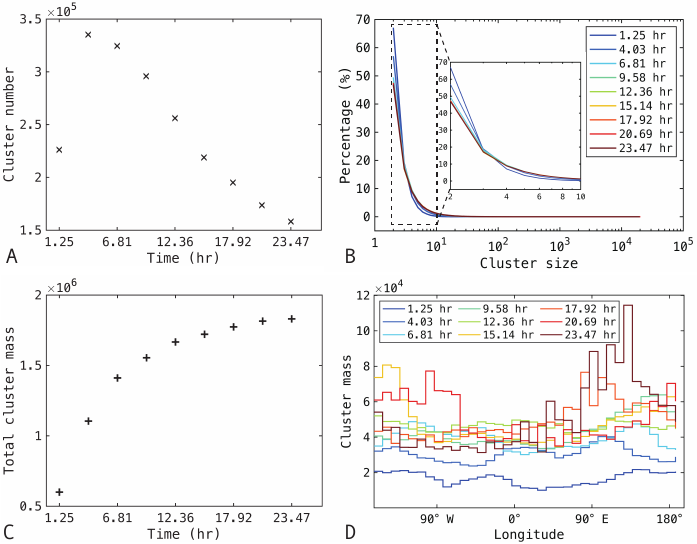}
	\caption{Statistical measure of aggregate growth in the DEM simulation involving 10 cm particles. A. The variation of cluster number with the simulated time. B. The percentage of  different sized clusters at different time points. The cluster size is presented by the quantity of particles it contains. The local zoom of cluster size between 2 to 10 is shown as the box inside. C. The total mass of clustered particles changing as a function of time, which means the cumulative mass of particles engaged in clusters. D. The distribution of cluster mass as a function of longitudes in different evolutionary stages. The cluster mass in panels C and D is normalized by the mass of an individual particle. An animation of the DEM simulation involving 10-cm debris particles is available, from which the statistical measure of panels A--D is abstracted. The animation lasts for 42 s and show the 24-hr evolutionary scenarios in global and local views. The right panel of the animation shows the magnified view of a 40-m squared cube (corresponding to the white box marked in the left panel) around the largest cluster, and the left panel provides a global top view around the asteroid. Note that 10 times of the original particle sizes are drawn for visual enhancement in the left panel. \label{f:clusterstat}}
\end{figure}

In the simulation involving $10$ cm particles, collisions are frequent in the first 4 hrs and a significant number of small clusters form initially, each containing a limited number of particles. After that, these small clusters continue to collide with each other or with dissociative debris particles, leading to mergence and growth. Figure~\ref{f:clusterstat} illustrates the collisional evolution of the debris cloud. As shown in Fig.~\ref{f:clusterstat}A, the number of clusters increases sharply during the first $4$ hrs, and the majority are small clusters. More than 96\% of the clusters formed in the first 4 hrs are composed of less than 10 particle monomers according to the statistics in Fig.~\ref{f:clusterstat}B. The cluster number peaks at 4.03 hr and then starts to decrease monotonously, however the total mass of all clustered particles shows a continuous increase and approximates a saturation that accounts for 67.08\% of the shedding mass (Fig.~\ref{f:clusterstat}C). The statistics show the accretion does not halt and the cluster number reduces because small clusters collide and merge, giving birth to fewer larger clusters. The percentage of clusters containing less than 10 particles decreases from 96.68\% at 4.03 hr to 88.20\% at 23.47 hr. At the end of the 24-hr simulation, large clusters that encompass over $10,000$ particles form. As more particles are involved in the clustering process, the collisions between particles (or that between clusters) become less frequent, causing a slowdown of the collisional accretion. Figure~\ref{f:clusterstat}C shows $24$ hrs is a meaningful long term to observe the growth of clusters out of the debris cloud generated by Section~\ref{sec:chrctcld}, i.e., at the end of the simulation the collisions are rare and the debris cloud enters a steady state. We checked the mass distribution of the clusters changing as a function of time and position. As shown in Fig.~\ref{f:clusterstat}D, the cluster mass in longitude intervals (each interval $10^\circ$) exhibits subtle variations in the first $12$ hrs. The histograms are generated at different times and marked with solid lines in different colors. The largely uniform distribution is broken in the last $12$ hrs because large clusters containing thousands of particles start to form, leading to a pronounced concentration in the mass distribution of clusters within the longitude regions where large clusters are present in real-time.

Global and local evolutionary scenarios of debris particles are recorded in the animation of Fig.~\ref{f:clusterstat} in the online article. Different from the ballistic propagation stage, we found that the collisions and contact effects of debris particles are predominant at the considered density of debris. Clusters appear first at the mass-concentrated regions and then are transported to the vicinity of the asteroid. After intensive mixing, the cluster mass is no longer concentrated in the longitude regions with low geopotential, instead, it expands as the propagation of the debris cloud. At the end of the simulation, the cloud is dilute and the ballistic motion of particles and clusters becomes dominant in the later period of the post-shedding evolution. Meanwhile, the growth of the clusters becomes much slower compared with the early stage.

\begin{figure}[h!]
	\centering
	\includegraphics[width=1.0\textwidth] {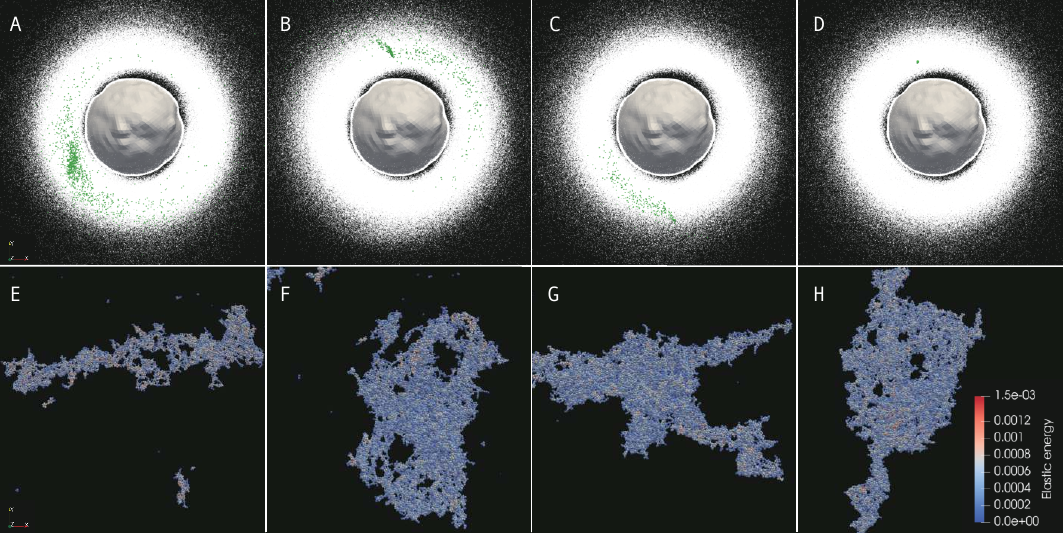}
	\caption{Snapshots of the debris cloud and the accretion process of the largest cluster in the DEM simulation involving $10$ cm particles. Panels A--D illustrate the collisional and aggregating process of debris particles, corresponding to the simulated time of 15.14 hr, 17.92 hr, 20.69 hr, and 23.47 hr respectively. The element particles of the largest cluster are rendered in green and 50 times the original sizes for visual enhancement. Panels E--H magnify the formation process of the largest cluster at the four time points (corresponding to panels A--D respectively) and the color denotes the inter-particle elastic energy with the unit of J. An animation of the $360^{\circ}$ scan of the cluster in panel H is available, showing the porous structure. The animation lasts for 16 s. \label{f:bigcluster}}
\end{figure}

Figure~\ref{f:bigcluster} illustrates the evolutionary process of the 10-cm debris cloud, including a global view and a zoomed view around the largest observed cluster. Snapshots are created at the time points of 15.14 hr, 17.92 hr, 20.69 hr, and 23.47 hr, showing the representative configurations of the debris cloud, the generation of the largest cluster from the cloud, and the morphological change of the cluster. The largest cluster formed in the $10$-cm simulation is composed of $19,704$ particles. Using the unique ID number of each particle of this cluster, we traced the aggregating process reversely. Fig.~\ref{f:bigcluster}A--Fig.~\ref{f:bigcluster}D depict the traced debris particles (in green) shrinking from a cloud to an aggregate (simulated time from 15.14 hr to 23.47 hr) in the complete cloud (in white). Traced particles are rendered 50 times of the original size to make an enhanced visualization. The mechanical environment generated by the irregular asteroid and the interaction between particles induce turbulent flow among the particles, which favors further collisional growth. Zooming the formation process of the largest cluster at the four corresponding moments, colliding, sticking, morphology changing, and merging of the fluffy-structured clusters are observed, as shown in Fig.~\ref{f:bigcluster}E--Fig.~\ref{f:bigcluster}H. The structure with voids inside possesses certain strength, which is verified by the measurement of the inter-particle elastic energy (defined as the summation of $k_\mathrm{n}\delta^2$). The animation of Fig.~\ref{f:bigcluster} in the online article provides a $360^{\circ}$ scan of the cluster with colors denoting the inter-particle elastic energy.

\begin{figure}[h!]
	\centering
	\includegraphics[width=1.0\textwidth] {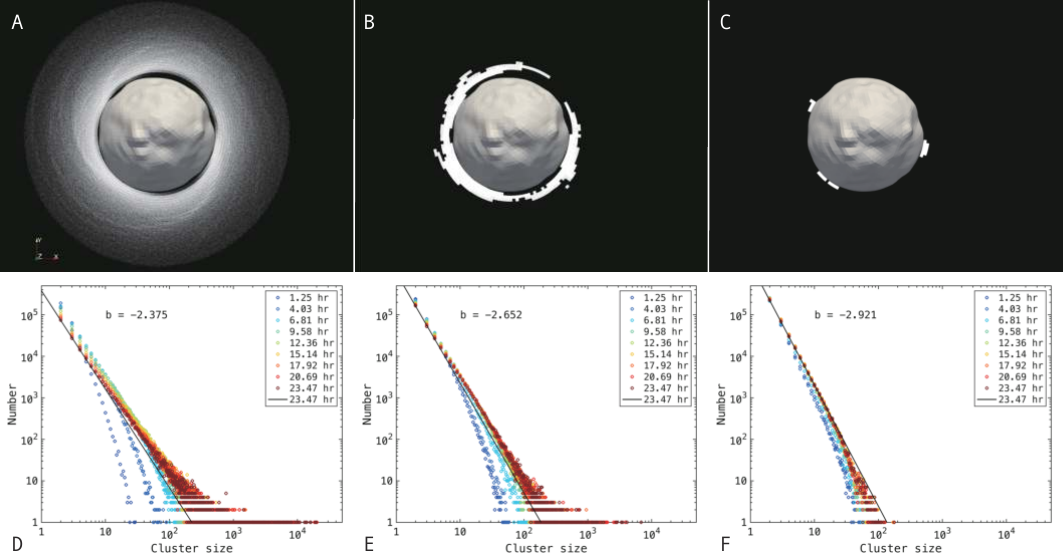}
	\caption{The initial distribution of debris particles and cluster size-number distribution shifting with time in the DEM simulations involving 10 cm, 5 cm, and 2 cm particles. Panels A and D correspond to the simulation involving 10 cm particles, panels B and E involving 5 cm particles, and panels C and F, 2 cm. The size of clusters is indicated by the number of particles they contain. Different colors denote different simulated moments. The cluster size-number distribution at 23.47 hr is fitted using power law with the power index $b$, as shown by the black lines. An animation of the propagation of 5 cm (the left panel of the animation) and 2 cm particles (the right panel) is available in the online article, whose first frame corresponds to panel B and C in this figure. The animation lasts for 18 s. \label{f:diffpt}}
\end{figure}

Simulations involving different-sized particles (different shedding mass) display differences in the development process of cluster formation in the simulated 24 hrs. Fig.~\ref{f:diffpt}A--Fig.~\ref{f:diffpt}C exhibit the initial distribution of debris particles in 10-cm-, 5-cm-, and 2-cm-debris cloud simulations, in which 2,728,342, 2,780,604, 2,729,293 particles are set up respectively as the initial condition. Considering the calculating capacity of millions of particles, only the densest regions are left and discretized into debris particles in 5-cm and 2-cm calculations. So the total shedding mass and the particle distribution of these three simulations are not identical, which should be considered when we compare and analyze the results. In the evolutionary process of 5-cm and 2-cm debris clouds, the particle clouds initially limited in the local region propagate to the whole ring rapidly, governed by the orbital manifolds (see the animation of Fig.~\ref{f:diffpt} in the online article). The extreme mass concentration is weakened gradually and the particle distribution w.r.t. the longitudes becomes more uniform. The collisional frequency in these two simulations is lower than that of the 10-cm simulation. The difference in the initial conditions should account for the result that no large clusters (containing more than 10,000 particles) formed in the simulated 24 hrs in 5-cm and 2-cm debris clouds. Fig.~\ref{f:diffpt}D--Fig.~\ref{f:diffpt}E depict the cluster size-number shifting as a function of time in log-log plots. Compared with the 10-cm-particle case, simulations with particle sizes of 5 cm and 2 cm exhibit lower aggregating efficiency in the simulated 24 hrs, even though the initial stages of small clusters growing rapidly are similar. During the identical simulated time of 24 hrs, clusters formed with magnitudes of only several thousands of particles and hundreds of particles in 5-cm and 2-cm cases respectively. Fitting the relation between cluster number $N_\mathrm{c}$ and cluster size $N_\mathrm{p}$ (the particle number a cluster contains) at 23.47 hr using the power law $N_\mathrm{c}\sim N_\mathrm{p}^b$, we get the power index $b=-2.375, -2.652, -2.921$ for the simulations of the three sizes. A reasonable guess is that the formation of large clusters (over 10,000 particles) from the debris clouds with small grains (radii below 10 cm) requires a longer time. Longer-duration simulations of a full debris cloud for 5 cm and 2 cm should be performed to obtain the scale of clusters when the aggregation approaches saturation, which, however, is beyond our computational capacity.

\section{Conclusion} \label{sec:concl}

This paper presented our study on the cumulative evolution of the debris cloud formed by a rotationally unstable asteroid. This short-term process is of particular interest for the initial growth of moonlets around a top-shaped asteroid, which is a typical type believed to be related to the slow spin-up effect of YORP. We derive a general formalism for the analysis of the dynamics of debris cloud originating from the asteroid surface. The cumulative effect of a continuous shedding event is calculated based on a ballistic model, which combines the irregular gravity, the solar radiation pressure, and collisions with the surface. The formalism deals with the macroscopic propagation of the debris cloud in the context of multiple forces. A hand-off algorithm from the ballistic model to DEM model is applied to define the initial conditions of the debris particles for calculating the particle-particle interactions. Large-scale DEM simulations are performed to examine the clustering behavior of the debris cloud as a function of the particle size. We believe the results are useful to identify the conditions that favor the birth of asteroid systems, and provide clues to the intrinsic mechanisms that decide the dynamical fate of a fast-rotating asteroid. 

We can draw the following conclusions by characterizing the macroscopic propagation of the debris cloud (Section~\ref{sec:chrctcld}): First, the debris cloud remains dense during and after the shedding process, and more turbulences are observed in the vicinity of the asteroid. Both factors contribute to increase the collision probability of the debris particles. Second, the size segregation is strong for particles below milimeter-size and becomes weak for particles over the centimeter level. For sub-kilometer fast-rotating asteroids, fine grains contribute little to the cumulation of a steady debris cloud and the collisional growth is dominated by large pieces of debris. Third, the cumulated debris cloud distributes unevenly and the concentrated regions exhibit a structural stability that is correlated to the local geopotential near the asteroid. The results reveal a common trend that the shed particles migrate from high geopotential to low geopotential via saltatory processes across the asteroid surface, which is a prevalent mechanism for asteroids approaching their critical spin limits. 

The conclusions from the DEM simulations of the post-shedding accretion process (Section~\ref{sec:accana}) are as follows: First, the simulation results confirm porous fluffy cluster structures can form shortly after a shedding event of observed magnitude as listed in Table~\ref{tab:shedobsev}. For a sketch of the life cycle of a cluster formation and growth, small clusters form abundantly from discrete particles at the initial stage, leading to a rapid population growth of small clusters; the initial stage is followed by a continuous slow accretion, during which the debris particles (or clusters) collide, scatter, and merge, leading to formation of larger clusters and morphological changes; as the collisional growth going, huge clusters (involving over 10,000 particles) form in 24 hrs and the surrounding debris cloud becomes dilute, which makes the inter-particle collisions less frequent; it slows down the collisional growth of the debris aggregates, and a saturation is observed at the end of the 24-hr simulation. Second, the clusters demonstrate porous fluffy structures possess certain strength and the adsorption capacity to dissociative debris particles colliding with it. Third, the collisional growth of debris particles in different sizes (2 cm, 5 cm and 10 cm) exhibits similar evolutionary stages as stated above. The population and size of the formed clusters follow the power law and the power slope drops with the particle size (under the same simulated time), i.e., the formation of large clusters requires longer time for smaller particles. 

\newpage
\section*{Acknowledgements}

The authors acknowledge Hera WG3 group for constructive conversations and useful discussions. Y.Y. acknowledges the financial support provided by the National Natural Science Foundation of China Grants No. 12272018.


\end{document}